\pdfoutput=1
\newif\iffull
\fulltrue
\newif\ifreview
\reviewfalse
\newif\iftodo
\todotrue

\ifreview
\documentclass[acmsmall,natbib,anonymous,review]{acmart}
\else
\documentclass[acmsmall,natbib]{acmart}
\fi
\usepackage{natbib}
\usepackage[nolist,printonlyused]{acronym}
\usepackage{listings}
\usepackage{booktabs}
\usepackage{siunitx}
\usepackage{multirow}

\AtBeginDocument{%
  \providecommand\BibTeX{{%
    \normalfont B\kern-0.5em{\scshape i\kern-0.25em b}\kern-0.8em\TeX}}}

\newcommand{\specialcell}[2][t]{%
  \begin{tabular}[#1]{@{}l@{}}#2\end{tabular}}

\def\secref#1{\S\ref{#1}}
\def\figref#1{Fig.~\ref{#1}}
\def\tabref#1{Tab.~\ref{#1}}

\def\eqref#1{Eq.~\ref{#1}}
\def\see#1{(\cf~#1)}

\def\eg{\emph{e.g.}}
\def\ie{\emph{i.e.}}
\def\cf{\emph{cf.}}

\def\c#1{\textbf{(C#1)}}

\def\cybert{\textsc{CySecBERT}}

\setcopyright{None}
\copyrightyear{2022}
\acmYear{2022}
\acmConference[Conference acronym 'XX]{}{2022}{Woodstock, NY}

\begin{document}

\title%
    [\cybert: A Cybersecurity Language Model]%
    {\cybert: A Domain-Adapted Language Model for the Cybersecurity Domain}

\author{Markus Bayer}
\authornote{Corresponding author}
\email{bayer@peasec.tu-darmstadt.de}
\orcid{0000-0002-2040-5609}
\affiliation{%
  \institution{PEASEC, Technical University of Darmstadt}
  \city{Darmstadt}
  \country{Germany}}

\author{Philipp Kuehn}
\email{kuehn@peasec.tu-darmstadt.de}
\orcid{0000-0002-1739-876X}
\affiliation{%
  \institution{PEASEC, Technical University of Darmstadt}
  \city{Darmstadt}
  \country{Germany}}

\author{Ramin Shanehsaz}
\email{r.shanehsaz@posteo.de}
\affiliation{%
  \institution{PEASEC, Technical University of Darmstadt}
  \city{Darmstadt}
  \country{Germany}}

\author{Christian Reuter}
\email{reuter@peasec.tu-darmstadt.de}
\orcid{0000-0003-1920-038X}
\affiliation{%
  \institution{PEASEC, Technical University of Darmstadt}
  \city{Darmstadt}
  \country{Germany}}
\begin{abstract}
The field of \acl*{cysec} is evolving fast.
Experts need to be informed about past, current and - in the best case - upcoming threats, because attacks are becoming more advanced, targets bigger and systems more complex.
As this cannot be addressed manually, \acl*{cysec} experts need to rely on machine learning techniques.
In the texutual domain, pre-trained language models like \acs*{bert} have shown to be helpful, by providing a good baseline for further fine-tuning.
However, due to the domain-knowledge and many technical terms in \acl*{cysec} general language models might miss the gist of textual information, hence doing more harm than good.
For this reason, we create a high-quality dataset and present a language model specifically tailored to the \acl*{cysec} domain, which can serve as a basic building block for cybersecurity systems that deal with natural language.
The model is compared with other models based on 15 different domain-dependent extrinsic and intrinsic tasks as well as general tasks from the SuperGLUE benchmark.
On the one hand, the results of the intrinsic tasks show that our model improves the internal representation space of words compared to the other models.
On the other hand, the extrinsic, domain-dependent tasks, consisting of sequence tagging and classification, show that the model is best in specific application scenarios, in contrast to the others.
Furthermore, we show that our approach against catastrophic forgetting works, as the model is able to retrieve the previously trained domain-independent knowledge.
The used dataset and trained model are made publicly available.
\end{abstract}

\maketitle

\section{Introduction}
\label{sec:introduction}
Cybersecurity evolved rapidely in recent years.
One of the offspring technologies is \ac{cti}.
Already with the first adoption in research and practise, which invited open discussions and collaboration on threat indicators, it became clear that the sheer volume of information could not be managed without automated support~\citep{hinchy_voice_2022}.
Meanwhile, the number of cyber attacks a day is steadily increasing~\citep{verizon}, while the COVID-19 pandamic\footnote{\url{https://enterprise.verizon.com/en-gb/resources/articles/analyzing-covid-19-data-breach-landscape/}} and the war in Ukraine \citep{eunCyberWarfareRussoUkrainian} intensifies this development.

Analyzing attacks is mostly manual work which includes reverse engineering and forensics.
The formats used to publish the resulting information differ as well.
They could be published in structured form as \ac{ioc}, or in natural language in blog posts and news articles.
The latter is the gist of \ac{cti}~\citep{soman2019blackhat, Wagner_Mahbub_Palomar_Abdallah_2019}.
Generally, this is a manual, labour intensive task in which experts extract actually relevant, evidence-based knowledge \citep{Tounsi_Rais_2018}.
This led to the idea of using \ac{nlp} to undertake this work and extract cyber threat information.

Recently, analyzing data by using neural network inspired language models has gained attention and has become an important part of modern \ac{nlp} systems~\citep{brownLanguageModelsAre2020}.
Particularly word embedding methods that use a sparse vector space to represent words are prominent instances~\citep{mikolovEfficientEstimationWord2013}.
In this context, models such as \ac{bert}~\citep{devlin2018bert} have become the standard basis models in all machine learning tasks that have natural language as input.
These models are already generally pre-trained and can be adapted to the task at hand, which is called fine-tuning.
Research has shown that the full potential of such models cannot be realized when applying them to domain-specific tasks~\citep{gururangan2020don, lee2020biobert, beltagy2019scibert, alsentzer2019publicly}.
This is intuitive because these models try to cover as many domains as possible, and specific domain knowledge is lost due to capacity constraints, especially in normal-sized models, or because the knowledge is not even included in the training data.
To gain domain-specific knowledge, these models can be further trained on domain-specific corpora to achieve even better results in this domain~\citep{lee2020biobert}.

Models trained on general domain corpora like Wikipedia have inherent problems with domain-specific purposes~\citep{mumtaz2020learning}.
They either have never seen domain-specific words, like new vulnerability names, or differentiate words with multiple meanings in multiple domains.
An example is the word \textit{virus} which might lead to a general model's understanding of a disease instead of some type of malware~\citep{mumtaz2020learning}.
This is troublesome for automated \ac{cti} since it misses fundamental threats, when searching for cyber threats.

In this paper, we propose \cybert, a word embedding model based on \ac{bert}~\citep{devlin2018bert} for analyzing cybersecurity texts.
Our aim is to  enable state of the art \ac{nlp} for the security domain and with that provide a model highly suitable for practical cybersecurity use cases and a solid base for further research in this field.
By evaluating our resulting model on different tasks~(\ie, intrinsic and extrinsic tasks) we ensure that it indeed enriches the cybersecurity domain as well as not forgetting too much of its preceding learnings.
In this study, we will pre-train a model on a thoroughly chosen cybersecurity corpus consisting of different datasets, such as scientific papers, Twitter, webpages, and the national vulnerability database.
A well-performing model for this use case may supersede a lot of manual work done by researchers.
Although there are well performing models for various specific purposes in this domain~\citep{lee2020catbert, rahali2021malbert}, the importance of a general cybersecurity model is undeniable.
The following contributions are achieved in this paper:
\begin{itemize}
    \item A pre-trained, general purpose cybersecurity language model based on \ac{bert} called \cybert~\c{1}.
    \item A sensibly chosen cybersecurity dataset containing all the data instances the model is trained on ~\c{2}.
    \item An evaluation of \cybert{} based on several tasks tailored to the cybersecurity domain, including intrinsic and extrinsic tasks, and general benchmark, to measure if and the degree to which the model forgets past knowledge~\c{3}.
    \item A comparison of our model to a related cybersecurity model, as well as the original \ac{bert} model and a discussion about its shortcomings and improvements~\c{4}.
\end{itemize}

%
%

%
\section{Related Work}
\label{sec:relatedwork}

This subsection gives an overview of relevant work on the topic of BERT models.
We outline models adapted to different domains that have emerged with the publication of \ac{bert}.
We also summarise work that already proposes BERT-like language models for the cybersecurity domain.
Finally, we indicate the research gap we are willing to fill.

%
%
%

\subsection{BERT Models in Different Domains} \label{section:domain_specific}

In various publications, the researchers were able to show that it is possible to achieve good results on domain-specific text corpora with pre-trained models such as BERT.
Of interest here is the method of \ac{dapt}~\citep{gururangan2020don}, which describes the process of training an already pre-trained language model on a domain-specific, domain-dependent dataset in the same way as the pre-training was done.
This differs from classical fine-tuning in that the model is not specialised for just one task, but serves as a building block for many tasks in the field. 
It is done in several other domains since the introduction of \ac{bert}~\citep{lee2020biobert,beltagy2019scibert,ranade2021cybert}. 
A prominent example is \textsc{BioBERT}, introduced by \citet{lee2020biobert}, where \ac{bert} was adapted to a biomedical corpus.
It was initialized with weights from \citet{devlin2018bert}'s \ac{bert} model and then pre-trained once again, this time with a large biomedical dataset, where the dataset was more than five times larger than \ac{bert}'s.
Using a subsequent fine-tuning process on three different biomedical text mining tasks, which are \acf{ner}, \acf{re}, and \acf{qa}, \citet{lee2020biobert} were able to largely outperform \ac{bert} and previous state-of-the-art models on these aforementioned tasks. 
Similar approaches present models that address other domains. 
\textsc{SciBERT}~\citep{beltagy2019scibert} for example focuses on scientific publications whereas \textsc{DA-RoBERTa}, introduced by~\citet{krieger2022domain} covers media bias. 
\citet{gururangan2020don} underpin our method of additional pre-training on \ac{bert}~by yielding good results applying this approach on \textsc{RoBERTa}~\citep{liu2019roberta}, a variant of \ac{bert} using the same transformer-based architecture.
In contrast to studies such as \textsc{BioBERT} by \citet{lee2020biobert}, in which only a single domain at a time is considered, \citet{gururangan2020don} covered a wide range of variations for pre-trained models to domains and tasks within those domains.

Similarly, researchers have also explored BERT models for the cybersecurity domain. 
For example, \citet{ranade2021cybert} propose a BERT model for this domain called CyBERT.
Although, the paper states that fine-tuning on \ac{bert} took place, in fact, the process is a further pre-training of \ac{bert} for the cybersecurity domain.
Fine-tuning is performed atop of this pre-trained cybersecurity model and is primarily used for application.
In general, they have a similar research goal as we do.

There are also further cybersecurity BERT models, which, however, are fine-tuned instead of continued pre-training in case of true \ac{dapt}, which makes them less suitable for other task of the cybersecurity domain.
\textsc{MalBERT}~\citep{rahali2021malbert} is a \ac{bert}-based model from the cybersecurity domain focusing on the detection of malicious software. 
Another security-related work is \textsc{CatBERT}, introduced by \citet{lee2020catbert}. 
They replaced some transformer blocks with adapters and fine-tuned the BERT model for the detection of phishing emails.
\citet{mendsaikhan2020identification} introduced a \ac{bert}-based Natural Language filter for identifying and classifying cyber threat-related information from publicly available information sources with high accuracy. 

\begin{table}
\centering
\begin{tabular}{lllll}

\toprule
Model / Paper & Domain / Use Case & Method & Model Base \\ 
\midrule
\textsc{BioBERT}~\citep{lee2020biobert} & biomedical & PT (+ FT) & \acs*{bert} \\
\textsc{SciBERT}~\citep{beltagy2019scibert} & scientific & PT (+ FT) & \acs*{bert} \\
\cite{gururangan2020don} & \specialcell{Papers (bio. +\\ CS), news, reviews} & PT (+ FT) & \textsc{RoBERTa} \\
\textsc{MalBERT}~\citep{rahali2021malbert} & malware & FT &  \specialcell{BERT,\\ \textsc{RoBERTa},\\DistilBERT} \\
\textsc{CatBERT}~\citep{lee2020catbert} &  \specialcell{phishing} & FT & DistilBERT \\
\textsc{ExBERT} & \specialcell{exploit prediction} & FT & \acs*{bert} \\
\cite{mendsaikhan2020identification} & CTI & FT & \acs*{bert} \\
\bottomrule
\end{tabular}
\caption{Overview over relevant existing BERT models for special domains. The method explains if the model was only fine-tuned (FT) or also pre-trained (PT).}
\label{tab:bertmodels}
\end{table}


An overview of the approaches with their domains and how they are trained can be found in Table \ref{tab:bertmodels}. 
These works are related to our work because the approach of adapting \ac{bert} to a specific domain is similar to our work and differs mainly in the target domain.
So, all in all, the different pre-trained \ac{bert} approaches can be used in our work as an orientation and also for comparing our results to theirs w.r.t. the performance.
Notwithstanding the fact that \ac{bert} has achieved great results in various domains, the full potentialities for the cybersecurity domain have yet to be exploited.


\subsection{Research Gap}

The research gap has led us to develop a model with the aim of achieving satisfactory performance for cybersecurity textual material in various tasks.
BERT has already been transferred to different domains, resulting in domain-specific models (\textsc{BioBERT} \citep{lee2020biobert}, \textsc{SciBERT} \citep{beltagy2019scibert}) and even specific domains in the cybersecurity domain, leading to models like \textsc{MalBERT} \citep{rahali2021malbert} or \textsc{CatBERT} \citep{lee2020catbert}.

As outlined in the introduction, there are a multitude of research problems in the field of cybersecurity based on the essential part of information extraction.
A solid method to address this can improve research in this field at a stroke.
Furthermore, it enables extensibility and additional layers can be applied on top of the model, such as CRF \citep{souza2019portuguese},~(Bi)LSTM, or both combined~\citep{jiang2019bert}.


\citet{ranade2021cybert} also addresses the delineated research gap to some extent.
Unfortunately, there was no juxtaposition with the results of \ac{bert} as the baseline but only a presentation of their model's outcome.
We compare our \cybert~with theirs, which is varied in the model training and the corpus \citep{ranade2021cybert}.
Furthermore, in delimitation to their work, we also evaluate a whole span of different cybersecurity tasks, ranging from classification to \ac{ner} and clustering tasks and we include the results of \ac{bert} for comparison reasons.
The latter results from the fact that we also take into account the phenomenon of catastrophic forgetting, where the pre-trained model forgets its already acquired knowledge in the new training phase, which has not been studied in other work in this area.
The similarity in both works results from the nature of the research task and also underlines the importance of the approach.
It is encouraging and important at the same time that there is such attention for this research gap.
Multiple works addressing a similar objective can be complementary and, therefore, expedite filling the gap in research.
Nonetheless, our work is distinguished from this paper at several points including the evaluation step, the applied data, and overall the extent of our work.

\section{Methodology}
\label{sec:methodology}
This section presents a short background on domain adaptive pre-training, including the planned training process, the dataset used to adapt our proposed language model to the cybersecurity domain, and the evaluation process.

\subsection{Domain Adaptive Pre-Training}
\label{subsec:methoddomainadaptivepretraining}

\Ac{dapt} of language models to a specific domain is a common method to achieve an advanced domain-specific language model~\see{\secref{sec:relatedwork}}.
It has been shown to increase model performance in several ways, from model performance on downstream tasks, and hence better evaluation results, to reduced training time for such tasks due to smaller datasets for the training process to achieve similar performance.
These prospects lead us to expect that the cybersecurity domain will benefit greatly from a domain-adapted, pre-trained language model for every possible task, \eg, \ac{ner} and relevance classification, to name a few.

We aim to adapt \ac{bert} to the cybersecurity domain based on domain specific text corpora~\see{\secref{subsec:methodtextcorpus}} \citep{gururangan2020don}.
Our \ac{dapt} pipeline is build with \emph{Huggingface}\footnote{\url{https://huggingface.co/}} and \emph{Weights and Biases}\footnote{\url{https://wandb.ai/}}.
The final domain-adapted pre-trained model is based on \verb!bert-base-uncased!.
Likewise, the text corpus is tokenized using the \verb!bert-base-uncased! model.
The training itself is done on the Lichtenberg Cluster\footnote{\url{http://www.hhlr.tu-darmstadt.de/}}.

During the training phase, we try to mitigate the problem of catastrophic forgetting~\citep{mccloskey1989catastrophic} by reducing the learning rate, the training steps, and the size of the dataset compared to \ac{bert} pre-training. 
In this way, the susceptibility to catastrophic forgetting should be greatly reduced because the new learning process is subordinated.
Nevertheless, we test whether the problem also occurs with the created model by evaluating it on a non-cybersecurity task. 
While we expect no improvements, we want to analyse to what extent the old knowledge has altered.

\subsection{Text Corpus}
\label{subsec:methodtextcorpus}

\stepcounter{footnote}
\newcounter{footnotesaver}\setcounter{footnotesaver}{\value{footnote}}
\footnotetext[\value{footnotesaver}]{Minimal or maximal token per entry.}

\begin{table}
\begin{tabular}{@{}lrrrrrr@{}}
\toprule
\#Tokens & {Min\footnotemark[\value{footnotesaver}]} & {Max\footnotemark[\value{footnotesaver}]} & {Sum} & {Median} & {Mean} & {Entries} \\
\midrule
Blogs    &   44  &  0.1M  &  169M  &   710  &  1.1k  &  151k \\
arXiv    &  533  &  0.7M  &  167M  &  8.2k  &  9.9k  &   16k \\
NVD      &    5  &  1.9k  &   12M  &    58  &    71  &  171k \\
Twitter  &    1  &   500  &  179M  &    39  &    45  &    4M \\ \addlinespace
Total    &    1  &  0.7M  &  528M  &    40  &   122  &  4.3M \\
\bottomrule
\end{tabular}
\caption{Statistics of the number of tokens based on the subset of our training dataset.}
\label{tab:datasetanalysis}
\end{table}

When creating the text corpus, we paid a lot of attention to the quality of the data, as this quality transfers to the model \citep{bayerSurveyDataAugmentation2022}. 
The text corpus is composed of different sub-corpora:
(i)~blog data,
(ii)~arXiv data,
(iii)~\ac{nvd} data, and
(iv)~Twitter data.
This decision is based on the kind of information, that can be found in either source and the fact, that the information is used in recent publications regarding machine learning.
Furthermore, they vary widely in their structure. 
While the NVD contains short, objective and precise language with semi-structured information, Twitter consists of short posts with objective, subjective, emotional, on- as well as off-topic, etc. content, arXiv encompasses long papers with highly educational language, and blog posts are typically longer articles with less formal language.

The blog posts build a solid foundation for different writing styles and practical information in information security, including vulnerability and exploit information, threat notifications~\citep{liao_acing_2016}, and foundational knowledge.
We crawled \num{38} different blogs, filtered duplicates and instances shorter than 300 characters\footnote{A randomly selected and manually inspected sample of short entries showed that lower length blog posts contained mostly advertisements or cookie notifications.}, resulting in over 151k blog posts.

Second, we use arXiv papers from the category \emph{Cryptography and Security}\footnote{For text extraction we used \href{https://github.com/pkubowicz/opendetex}{opendetex} for papers in tex format or \href{https://textract.readthedocs.io/en/stable/}{textract} for papers in pdf format.}~\citep{lee2020biobert}.
Due to errors during the text extraction process, we ignored papers with lower length than \num{3000} characters, resulting in over 16k papers.

Third, we use vulnerability descriptions of the \ac{nvd}~\citep{kuehn_ovana_2021,dong2019towards}.
Experts curate those texts\footnote{\url{https://www.cve.org/ResourcesSupport/FAQs\#pc_cve_recordscve_record_descriptions_created}}, so they need no further processing.
Hence, we do neither filter nor pre-process these information.

Lastly we use Twitter as information source~\citep{chen_using_2019,sabottke_vulnerability_2015,riebe_cysecalert_2021}.
We crawled datasets with the following keywords:
\begin{itemize}
    \item \emph{infosec OR security OR threat OR vulnerability OR cyber OR cybersec OR infrasec OR netsec OR hacking OR siem OR soc OR offsec OR osing OR bugbounty}
\end{itemize}
Additionally, we also crawled dedicated datasets of data breaches, as, for example, the Microsoft Exchange Server Data Breach.
Overall, we managed to crawl nearly 4M tweets with over 179M tokens in total.

A summary of all datasets is depicted in \tabref{tab:datasetanalysis}.

\subsection{Setup}
\label{subsec:evalsetup}
\begin{figure}
    \includegraphics[width=0.7\textwidth]{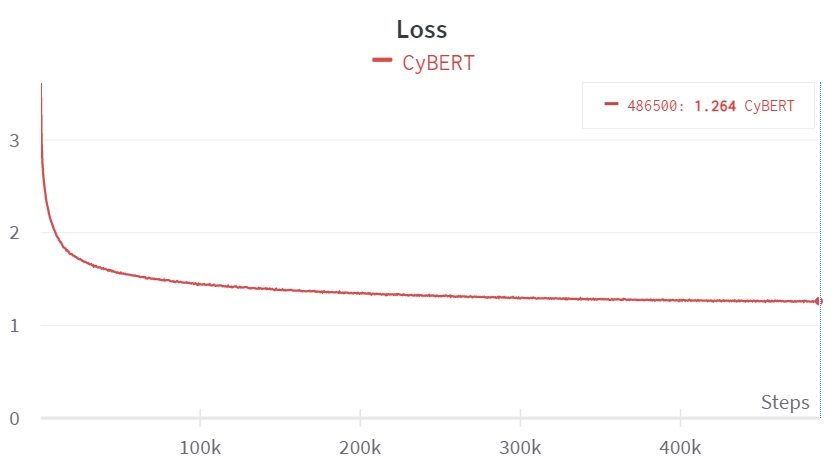}
    \caption{Training loss of the \cybert\ model.}
    \label{fig:loss}
\end{figure}

While the authors of \ac{bert} trained the model with a learning rate of \num{1e-4} for about \num{40} epochs, we adapted \ac{bert} to the cybersecurity domain with a learning rate of \num{2e-5} and \num{30} epochs on a dataset that is 10\% of the size of the original BERT dataset.
We trained our proposed model with a batch size of 64 on 4 Nvidia~Tesla~V100 GPUs. 
For the other hyperparameters, we followed the original BERT work, \ie, we used a weight decay of \num{0.01}, a dropout rate of \num{0.1}, \num{10000} warm-up steps, and ADAM as the optimisation algorithm~\citep{kingma_adam_2017}.
The training loss of the \cybert\ model can be seen in \figref{fig:loss}. 
It shows that the loss decreases logarithmically and only improves very slowly after 300k steps.

\section{Evaluation}
\label{sec:evaluation}

In this section, the evaluation process and the corresponding results are presented in detail.
We cover a short description of the evaluation tasks in \secref{subsec:methodevaluation}. 
After this the presentation and interpretation of the results follow in \secref{subsec:evalresults}.

\subsection{Experiments and Tasks}
\label{subsec:methodevaluation}

Since our goal is to publish a model that is highly usable for the cybersecurity domain, we evaluate it against the current standard method of the domain (BERT) and another cybersecurity language model (CyBERT from \citet{ranade2021cybert}).
We use different types of tasks, \ie~intrinsic and extrinsic evaluation tasks, which are reasonably chosen in the field of cybersecurity.
While the extrinsic tasks measure how well the trained model performs on downstream tasks, \ie~measure real-world application, the intrinsic tasks measure the model itself without any kind of additional classifier, \eg~by measuring the representations of the model and showing an overall fit to the domain.

As intrinsic tasks we use a word similarity task using parts of the dataset by \citet{mumtaz2020learning} and
a Twitter dataset for clustering evaluation. 
The clustering dataset is based on a random sample of Log4j Twitter posts.
For this task, the posts are converted into latent representations with the different \ac{bert} models. 
The latent representation consists of the concatenation of the last four layers of the model output and then the mean values over all words in the post. 
A KMeans clustering algorithm with k-values from 5 to 9 is executed on the gathered and transformed posts.
The evaluation scores are measured with the Silhouette Coefficient~(the higher the better).
This is an internal clustering metric that analyses the clusters created and does not require gold labels.
This is important because there can be many solutions and gold labels can be misleading in the case of clustering \citep{bayerInformationOverloadCrisis2021}.
The cybersecurity word similarity dataset consists of words with their similar words, all from the field of cybersecurity. 
The dataset is an extension of the public cybersecurity word similarity dataset from \citet{mumtaz2020learning} and contains over 300 word pairs.
Normally, the evaluation is based on the cosine similarity of the word embeddings when static embeddings are used.
However, BERT is context-dependent, so the cosine similarity of BERT-encoded words is not a good metric for scoring.
For this reason, we developed a novel word similarity evaluation method that asks the model to predict whether two given words are similar in a zero-shot learning setting.
Similar to the works on zero-shot learning, we create a meaningful cloze task for the model consisting of a sentence with a masked word that the model fills in, which is implicitly the answer to the similarity question.
The task is written in the following way: 

\indent ``\textit{Are $word_1$~and $word_2$~similar? \textbf{[MASK]}''}, where \textbf{[MASK]}~can be either ``Yes'' or ``No'', which represent the masked words that the model has to fill.

Example: \textit{``Are \textbf{virus}~and \textbf{malware} similar? \textbf{[MASK]}''}

Since we also want to show that the model does not only predict that every word is similar to every other word, we also randomly take word pairs from the dataset that are not similar and add them to the evaluation. 
This dataset is then used and an F1 score is calculated for the model's predictions.

As part of the extrinsic tasks, we use two cybersecurity classification tasks from \citet{riebe_cysecalert_2021} and \citet{bayerMultiLevelFineTuningData2022}. 
In the first task, the classifier has to decide whether a Twitter post is related to the field of cybersecurity, and in the second task, it has to predict whether a post might be relevant to experts in the field during a major cybersecurity event. 
Furthermore, we use the sequence tagging dataset by \citet{kuehn_ovana_2021}.
Sequence tagging is the task of finding specific words in a text and means that each word in a text is tagged, often as IOB: Inside, outside, and beginning, referring to the specific words being searched for.
Specifically, the dataset consists of several NER tasks for predicting relevant details of NVD descriptions.
We chose the task of predicting the name and version of the software and the attack vector. 
These tasks were chosen for their different performances in \citet{kuehn_ovana_2021}'s work, in order to analyse how well the models perform on different difficulties of the tasks.

Furthermore, we evaluate \cybert\ and \ac{bert} on the SuperGLUE benchmark~\citep{wang_superglue_2020}.
This is a common NLP benchmark used in this work to identify signs of catastrophic forgetting~\cite{mccloskey1989catastrophic}.
We assume that our cybersecurity model is not able to achieve a better or even equivalent score, but still think that the scores should not be too bad, as this would indicate that some basic knowledge would have been forgotten during the domain training phase.

Due to the scatter of results, all extrinsic experiments were performed five times, and the mean values as well as the standard deviation are given if they are informative. 

\subsection{Results}
\label{subsec:evalresults}
As stated in \secref{subsec:methodevaluation}, we aim to evaluate \cybert, as well as \ac{bert} and CyBERT~\citep{ranade2021cybert}, on different tasks, mainly settled in the \ac{cysec} domain, both intrinsic and extrinsic evaluation tasks.
Additionally, we run the SuperGLUE task to test our model for catastrophic forgetting.

\subsubsection{Intrinsic Tasks}
To measure the representation quality of the model, we evaluate it using intrinsic tasks from \ac{cysec}: clustering and word similarity.

\paragraph{Clustering} The results of the clustering task are given in Table \ref{tab:clustering}. 
While the CyBERT model of~\citet{ranade2021cybert}, is only better than the baseline when forming 5-7 clusters, our approach is better in every constellation, according to the Silhouette Score.
\cybert\ actually outperforms \citet{ranade2021cybert}'s cybersecurity model by a considerable margin for each number of clusters, with consistent improvements ranging from +0.002 to +0.059 points.
Our model shows the highest improvement when 9 clusters are formed.
On the one hand, these results show that we can obtain more coherent clusters thanks to our trained language model. 
On the other hand, from a more general perspective, the results show that the model is better able to represent the given instances in a meaningful latent space.

Nevertheless, even better results can be expected if we use an approach like SentenceBERT by \citet{reimersSentenceBERTSentenceEmbeddings2019} for our model, as they prove to be much better for representing complete documents, like tweets in our case.

\paragraph{Word Similarity} The word similarity task results are shown in \tabref{tab:intrinsic}. 
The baseline model has the worst performance with a F1-score of 0.44.
This is to be expected, as most domain-specific words were not or only very rarely included in the standard BERT training.
Our \cybert model is clearly superior to the other two approaches, which is very interesting to see and confirms the previous intrinsic results.
However, we would like to point out that this task is different from other word similarity tasks and does not reflect the word similarities directly through the word representations, but by questioning the model in a cloze fashion (see \ref{subsec:methodevaluation}).

\begin{table}[!htb]
\begin{minipage}{.4\linewidth}
    \centering

    \medskip
\begin{tabular}{llll}
    \toprule
    \# Clusters & BERT  & \citep{ranade2021cybert} & \cybert           \\ \hline
    5           & 0.114 & 0.141  & \textbf{0.143} \\
    6           & 0.115 & 0.124  & \textbf{0.150} \\
    7           & 0.118 & 0.133  & \textbf{0.167} \\
    8           & 0.125 & 0.117  & \textbf{0.163} \\
    9           & 0.130 & 0.113  & \textbf{0.172} \\
    \bottomrule
    \end{tabular}
    \caption{Silhouette Score of the first intrinsic task, clustering the data of a Log4J dataset. The best values are marked.}
    \label{tab:clustering}
\end{minipage}\hfill
\begin{minipage}{.4\linewidth}
    \centering
    \medskip
\begin{tabular}{@{} lll @{}}
    \toprule
        Tasks            & Word Similarity \\ \midrule
        \ac{bert}        &   0.4382              \\
    \citep{ranade2021cybert} &   0.4861             \\
        \cybert          &   \textbf{0.6382}              \\
    \bottomrule
    \end{tabular}
    \caption{Overview of the results of the word similarity task where the scores are indicated by the F1 score.}
    \label{tab:intrinsic}
\end{minipage}
\end{table}

\subsubsection{Extrinsic Tasks}
Now that we have shown that the model produces meaningful representations of cybersecurity words and data, we want to check whether our model is also more suitable for real-world applications, i.e. for extrinsic tasks, the so-called downstream tasks of machine learning.
The tasks that we have chosen for the cybersecurity domain are (i) \ac{ner} (ii) general relevance classification, and (iii) CTI classification .

\paragraph{NER} The results of the \ac{ner} task are shown in \tabref{tab:ner_evaluation}.
While we can see that the basic BERT model and the CyBERT model of \citet{ranade2021cybert} are more similar, e.g. in terms of software naming (SN), our model consistently outperforms both.
Only in the tagging of the software version (SV) does the CyBERT model of \citet{ranade2021cybert} perform significantly better than the baseline BERT model, while our model nevertheless improves this result. 
One can speculate that the CyBERT training data from \citet{ranade2021cybert} contained software versions at a higher frequency than the normal BERT data. 
However, the CyBERT model deteriorates the results for software names (SN), which could indicate that either a large number of software names are missing in high frequency in their dataset or they have been neglected due to errors in the training process.
The highest improvements of our model are seen in attack complexity (AC), with 0.0136 points better than the CyBERT model of \citet{ranade2021cybert}.
Nevertheless, the results of this particular task are not very satisfactory, which has already been discussed by \citet{kuehn_ovana_2021} and is related to the problem of too little data in this task. 

\begin{table}[]
    \centering
    \begin{tabular}{@{}l rr rrrrr@{}}
    \toprule
        \acs*{cvss} \acs*{ner} & SV     & SN     & AC     \\ \midrule
        \ac{bert}              & 0.9247 (0.0064) & 0.8837 (0.0037) & 0.3323 (0.0135) \\
      \citep{ranade2021cybert} & 0.9298 (0.0019) & 0.8834 (0.0029) & 0.3336 (0.0214) \\
        \cybert                & \textbf{0.9302} (0.0066) & \textbf{0.8871} (0.0025) & \textbf{0.3472} (0.0116) \\
    \bottomrule
    \end{tabular}
    \caption{\Acl*{ner} score based on tagged software versions (SV), software names (SN), and attack complexities (AC) of NVD descriptions. The results are given as F1 scores and the best values are marked.}
    \label{tab:eval_extrinsic_ner}
    \label{tab:ner_evaluation}
\end{table}

\paragraph{Relevance Classification (CySecAlert)} 
In the first classification task of our experiments, the models are trained to predict whether a Twitter post is related to the cybersecurity domain (see \tabref{tab:classificationresults}).
This can be considered a general cybersecurity task, as the model only needs to identify cybersecurity-related words.
In this context, it is particularly interesting to see that the \citet{ranade2021cybert} model is worse than the basic BERT model.
Our model significantly improves the base model and the CyBERT model by 0.0104 and 0.0236 points in the F1 score, respectively.
All models have a relatively low standard deviation, indicating that the fine-tuning process is stable across runs.

\paragraph{Specialised CTI Classification (MS Exchange)} 
The second classification task is about finding specialised CTI where very specific words are needed to classify the instances.
The results are also presented in \tabref{tab:classificationresults}.
Surprisingly, unlike the other tasks, the CyBERT model of \citet{ranade2021cybert} is able to improve the baseline, showing that while it does not improve the more general tasks, it could be beneficial in more specific tasks.
Here, there is a high improvement of our model compared to the baseline observation (+0.027), which we expected since this task focuses on very domain-dependent language and specific words.
Moreover, although the CyBERT model of \citet{ranade2021cybert} is advantageous for this task, our model still improves the results significantly by +0.0103 F1 points.
Here we also see that our model has a significantly lower standard deviation than the other two models, which again indicates a very stable training process.

\begin{table}[]
\centering
\begin{tabular}{lll}
\toprule
       & MS Exchange              & CySecAlert               \\ \hline
BERT   & 0.8599 (0.0193)          & 0.8779 (0.0084)          \\
\citep{ranade2021cybert} & 0.8766 (0.0153)          & 0.8647 (0.0095)          \\
\cybert   & \textbf{0.8869 (0.0026)} & \textbf{0.8883 (0.0064)} \\
\bottomrule
\end{tabular}
\caption{Classification results of the MS Exchange and CySecAlert dataset, given as F1 scores. The best values are marked.}
\label{tab:classificationresults}
\end{table}

\paragraph{Catastrophic Forgetting}

In the last part of our evaluation we address the problem of catastrophic forgetting.
To this end, we evaluate our model with the SuperGLUE benchmark to see if the model degrades the results too much, which would indicate that the model has forgotten the initial knowledge acquired in the BERT training phase.
The results of this task and a comparison to the BERT model is shown in Table \ref{tab:superglue_evaluation}.
As expected, we can see that our model reduces almost every task outcome. 
Nevertheless, the worse results do not indicate catastrophic forgetting, as the differences are still relatively small, with a mean drop of about -0.05 points.
This shows that although the model has lost some of its knowledge, it still has most of it left.
It is interesting to see that the cb task has even increased and the result of the boolq task has remained almost the same.

\begin{table*}
\begin{tabular}{lrrrrrrrrr}
\toprule
{}             &  record &     rte &     wic &     wsc &   boolq &      cb &   copa &  multirc &  total\_score \\
\midrule
BERT           &  0.6416 &  0.5949 &  0.6476 &  0.5538 &  0.6760 &  0.3704 &  0.606 &   0.4067 &     0.600994 \\ 
\cybert    &  0.6137 &  0.5545 &  0.5887 &  0.5404 &  0.6752 &  0.5551 &  0.486 &   0.3915 &     0.546831 \\
\bottomrule
\end{tabular}
\caption{Results of the SuperGLUE benchmark, each indicated in the evaluation metric proposed in the benchmark.}
\label{tab:superglue_evaluation}
\end{table*}

\paragraph{Conclusion}
In our evaluation, we have shown that the model developed in this work has very good cybersecurity capabilities.
The tasks showed that the \cybert~model is able to outperform the BERT baseline and the CyBERT model by \citet{ranade2021cybert} consistently across all cybersecurity tasks.
We evaluated these models on intrinsic cybersecurity tasks where we showed how well the models represent documents and words in latent space. 
These tasks assess the fundamental quality of the language model.
In addition, we evaluated the three models using extrinsic cybersecurity tasks that demonstrate the practicality of the model in most real-world application contexts.
Our model improves the results of these tasks by up to 0.027 F1 points compared to the other two models and achieves its highest improvement on an in-depth cybersecurity task where very specific language differences have to be considered.
In addition, we also analysed the phenomenon of catastrophic forgetting by evaluating our model on standard NLP tasks.
Although there is a deterioration in performance in these tasks, it is only within the expected range of decline.
We can say with confidence that our model is capable of handling a wide range of cybersecurity tasks while retaining the original language modelling knowledge.

\section{Discussion, Conclusion, and Outlook}
\label{sec:conclusion}

In this work we propose a novel state of the art cybersecurity language model based on BERT \citep{devlin2018bert}. 
We perform \ac{dapt} on this model with a sensibly chosen cybersecurity corpus.
The corpus consists of a variety of source data structures, such as blogs, paper, as well as Twitter data.
The data and frequencies in the sources were selected to be appropriate for cybersecurity research and practice. 
Furthermore, the size of the dataset and the structure of the training process were chosen to prevent catastrophic forgetting on the one hand, and on the other hand, so that the model learns enough to contribute to the general field and specific niches of cybersecurity.
We explore this through a thorough evaluation of various tasks and in comparison to the BERT baseline as well as the current state of the art of cybersecurity language models.
First, we evaluate the models on two intrinsic tasks, where we show that the quality of our model improves in terms of the learned representation space, i.e. how well the cybersecurity-specific instances (words and texts) can be distinguished from each other.
Table \ref{tab:clustering} and \ref{tab:intrinsic} show the substantial performance increases by our model.
Second, we evaluated the model together with the other two models for cybersecurity-specific classification and NER tasks to show the usefulness and practicality of the model in application contexts.
Our model outperforms the other models in every task, which can be seen in Table \ref{tab:eval_extrinsic_ner} and \ref{tab:classificationresults}.
The greatest improvement is observed in the special CTI classification dataset, suggesting that the model is particularly beneficial when dealing with very specific cybersecurity language that a normal BERT model did not have in the training dataset.
Our evaluation concludes with a focus on catastrophic forgetting by assessing the performance of our models against a general NLP benchmark.
While these results (table \ref{tab:superglue_evaluation}) show that our model does indeed degrade the results, they also show that, as intended, there is no catastrophic forgetting and that the final model has combined much of its original knowledge with the new knowledge about cybersecurity.

While we are aware that the current state of the art on research in language modeling and NLP generally tends to focus on larger language models, like GPT-3 by \citet{brownLanguageModelsAre2020}, we have chosen the BERT model on purpose.
Most of the cybersecurity research and especially practice does not have the necessary resources to apply large language models. 
In most cases, the BERT model can still be considered the standard model in such ML application contexts as the cybersecurity domain. 
In this way, our work benefits most for the research landscape and practice.

\paragraph{Practical and Theoretical Implications}
Our work contributes to research and pratice through a novel, state-of-the-art cybersecurity model called \cybert, which is published. 
We also publish the associated dataset so that it can contribute to further work.
Thus, our work has several implications for practice and research:

\textbf{A novel, state-of-the-art language model for cybersecurity that is useful for various tasks.} 
With the research around the model, we aimed to find a solution to increase the performance of machine learning in as many cybersecurity language tasks as possible.
Our model provides utility for a large number of tasks, as it can be estimated based on the success in extrinsic task scores as well as inferred from intrinsic task scores, which show that the representation space is better for the domain-dependent language with our model.
With the release of the model, we are paving the way for better cybersecurity tools, as practitioners can easily use the new model in existing pipelines, for example, in alert aggregation \citep{landauerDealingSecurityAlert2022}, detection of phishing websites \citep{xiangCANTINAFeatureRichMachine2011,yangPhishingWebsiteDetection2019}, or even maleware detection \citep{salemMaatAutomaticallyAnalyzing2021}.
The better tools will then also be the result of new research derived from the model and will even improve the results in various tasks by incorporating further research ideas. 
This can be done on a smaller scale, where the model is not the focus but serves as a foundation on which further techniques such as data augmentation, meaningful data selection, few-shot learning or specific applications are built.
But it can also be done on a larger scale where the model is the subject of research, for example by analysing its results in explainable AI approaches.

\textbf{A sensibly chosen cybersecurity dataset containing most relevant sources.}
When creating the dataset, care was taken to include many different sources so that the model can be used for a wide range of cybersecurity tasks.
The publication of this dataset can be used in further work analysing the content, e.g. if there is a bias in the data.
The dataset can also serve as a basis for training other language models.
Although we have deliberately chosen this size of dataset for BERT training to prevent catastrophic forgetting, it might be useful to expand the dataset, which can easily be done by collecting more data sources that we have already selected.

\paragraph{Ethical Considerations}
We would like to emphasise that we did not explicitly focus on and analyse social biases in the data or the resulting model.
While this may not be so damaging for most application contexts, there are certainly applications that rely heavily on these biases, and including any kind of discrimination can have serious consequences.
As authors, we would like to express our warnings regarding the use of the model in such contexts.
Nonetheless, we aim for an open source mentality, seeing the great impact it can have, and therefore transfer the thinking to the user of the model, drawing on the many previous discussions in the open source community. 

\begin{acks}
We thank all anonymous reviewers of this work.
This research work has been funded by the German Federal Ministry of Education and Research and the Hessian Ministry of Higher Education, Research, Science and the Arts within their joint support of the National Research Center for Applied Cybersecurity ATHENE
and
by the German Federal Ministry for Education and Research~(BMBF) in the project CYWARN~(13N15407).
The calculations for this research were conducted on the Lichtenberg high performance computer of the TU Darmstadt.

\end{acks}

\bibliographystyle{ACM-Reference-Format}
\bibliography{bibliography}

\begin{acronym}
\begin{acronym}
\acro{ac}[AC]{Attack Complexity}
\acro{albert}[ALBERT]{A Lite BERT}
\acro{av}[AV]{Attack Vector}
\acro{bert}[BERT]{bidirectional encoder representations from transformers}
\acro{bi-lstm}[Bi-LSTM]{Bi-directional long short term memory}

\acro{ioc}[IoC]{indicator of compromise}
\acro{cti}[CTI]{cyber threat intelligence}

\acro{cbow}[CBOW]{continuous bag-of-words}
\acro{cve}[CVE]{Common Vulnerabilities and Exposures}
\acro{cvss}[CVSS]{Common Vulnerability Scoring System}

\acro{iot}[IoT]{Internet of Things}
\acro{jsonl}[JSONL]{JSON Lines}
\acro{kb}[KB]{knowledge base}
\acro{lstm}[LSTM]{Long short-term memory}

\acro{mee}[MEE]{Malware Entity Extractor}
\acro{ml}[ML]{machine learning}
\acro{mlm}[MLM]{masked language modeling}
\acro{dapt}[DAPT]{domain-adaptive pre-training}
\acro{ner}[NER]{named entity recognition}
\acro{nlp}[NLP]{natural language processing}
\acro{nltk}[NLTK]{Natural Language Toolkit}
\acro{nsp}[NSP]{Next Sentence Prediction}
\acro{nvd}[NVD]{national vulnerability database}

\acro{ovana}[OVANA]{Overt Vulnerability source ANAlysis}
\acro{qa}[QA]{Question Answering}
\acro{re}[RE]{Relation Extraction}
\acro{roberta}[\textsc{RoBERTa}]{Robustly optimized BERT approach}
\acro{sk}[SK]{skip-gram}

\acro{sn}[SN]{Software Name}
\acro{sv}[SV]{Software Version}
\acro{cysec}[CySec]{cybersecurity}
\end{acronym}
\end{acronym}

\end{document}